\documentclass[sigchi]{acmart}

\renewcommand\footnotetextcopyrightpermission[1]{} 
\usepackage{booktabs} 
\usepackage{listings}
\definecolor{myblue}{rgb}{0.12, 0.46, 0.7}

\setcopyright{rightsretained}

\acmDOI{10.475/123_4}

\acmISBN{123-4567-24-567/08/06}

\acmConference[UIST'18]{ACM UIST conference}{October 2018}{Berlin, Germany}
\acmYear{2018}
\copyrightyear{2018}

\acmPrice{15.00}

\begin{document}
\title{A Computational Method for Evaluating UI Patterns}

\author[Doosti et al.]{Bardia Doosti$^1$ \qquad Tao Dong$^2$ \qquad Biplab Deka$^3$ \qquad Jeffrey Nichols$^2$}
\affiliation{
       \begin{tabular}{ccc}
       $^1$Indiana University Bloomington & $^2$ Google Inc. & $^3$University of Illinois at Urbana-Champaign\\
       Bloomington, IN & Mountain View, CA & Champaign, IL\\
       bdoosti@indiana.edu & \{taodong, jwnichols\}@google.com & deka2@illinois.edu
       \end{tabular}
       }


\begin{abstract}
UI design languages, such as Google's Material Design, make applications both easier to develop and easier to learn by providing a set of standard UI components. Nonetheless, it is hard to assess the impact of design languages in the wild. Moreover, designers often get stranded by strong-opinionated debates around the merit of certain UI components, such as the Floating Action Button and the Navigation Drawer. To address these challenges, this short paper introduces a method for measuring the impact of design languages and informing design debates through analyzing a dataset consisting of view hierarchies, screenshots, and app metadata for more than 9,000 mobile apps. Our data analysis shows that use of Material Design is positively correlated to app ratings, and to some extent, also the number of installs. Furthermore, we show that use of UI components vary by app category, suggesting a more nuanced view needed in design debates.
\end{abstract}

%
%



\maketitle

\begin{figure*}[t]
    \centering
    \includegraphics[scale=0.3]{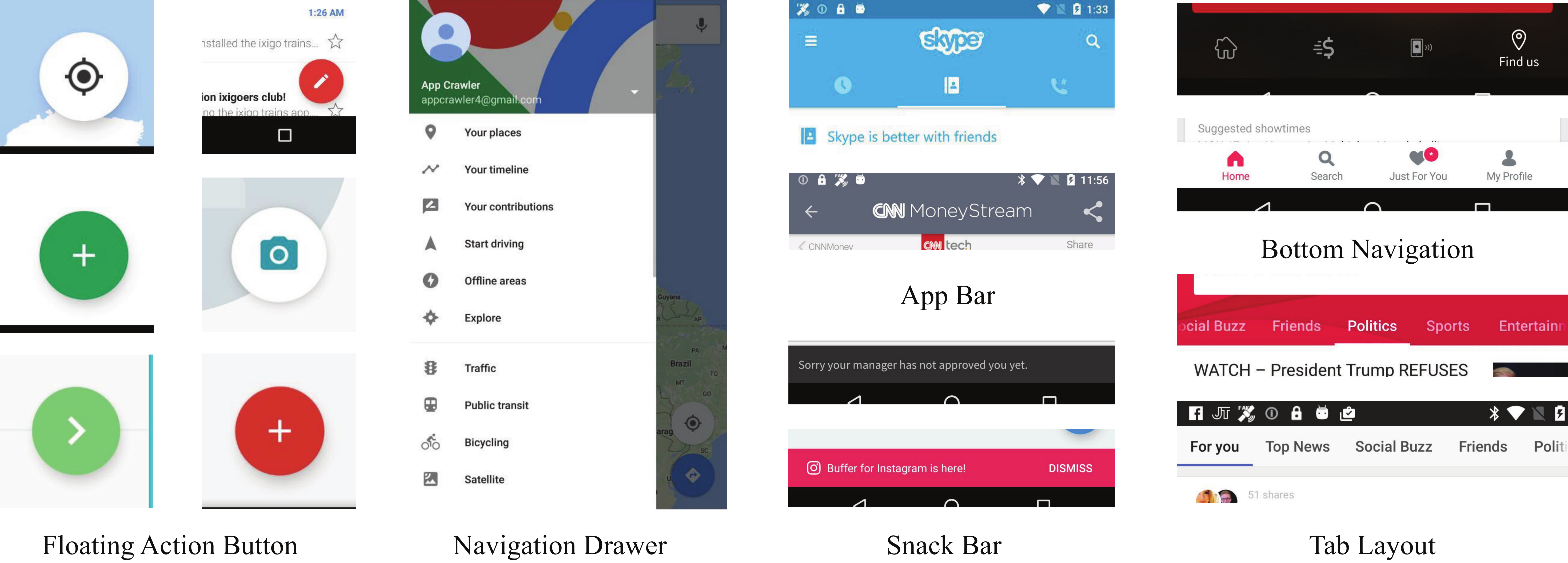}
    \caption{Some of the Material Design cropped components}
    \label{fig:cropped}
\end{figure*}

\section{Introduction}
Pattern languages have been long used in Human-Computer Interaction (HCI) for distilling and communicating design knowledge~\cite{Borchers:2000, dearden2006pattern}. According to Christopher Alexander~\cite{alexander1977pattern}, who introduced pattern-based design to architecture, ``each pattern describes a problem which occurs over and over again in our environment, and then describes the core of the solution to that problem, in such a way that you can use this solution a million times over, without ever doing it the same way twice.''  

HCI researchers and design practitioners have documented and introduced pattern languages for general UI design (e.g., \cite{tidwell2010designing} and \cite{neil2014mobile}) as well as a wide variety of application domains, such as learning management systems~\cite{avgeriou2003towards}, ubiquitous computing~\cite{landay2003design}, information retrieval~\cite{wania2009pattern} and many more. Nonetheless, as Dearden and Finlay point out, there have been relatively few evaluations of how useful pattern languages are in user interface design~\cite{dearden2006pattern}. Since Dearden and Finlay published their critical review, more evaluations have been done on pattern languages in HCI (e.g., \cite{wania2009pattern}, \cite{chung2004development} and \cite{saponas2006impact}). But these evaluations are usually limited in at least one of several ways. First, the pattern languages in those evaluations were often developed in an academic research setting. Few have been applied to real world applications. Second, the evaluations were usually done in lab settings, and hence lacked ecological validity. Last, those evaluations were done at a very small scale (i.e., applying a pattern language to either one or no more than a handful of systems). As a result of the limitations of how the field has evaluated pattern languages, we know little about whether pattern languages in HCI are fulfilling the promise of Alexander--providing design solutions that can be reused ``a million times over.'' 

The recent success of commercial UI design pattern languages offers a rare opportunity for us to evaluate the usefulness of pattern languages in HCI at scale and in the wild. In particular, Material Design\footnote{Material Design. \url{https://material.io/guidelines/}}, a UI design pattern language introduced by Google in 2014, seems to have been widely adopted by developers who build applications for Google's Android operating system. \textit{How can we understand the impact of a pattern language in one of the largest computing ecosystems in the world?} This is the first research question we seek to answer in this short paper.

In addition to developing a method for measuring a pattern language's overall impact, we also want to address questions about how and where certain patterns should be used when they get applied to new use cases. For Material Design, few patterns have been more controversial, yet at the same time iconic, than the Floating Action Button (aka, FAB) and the Navigation Drawer (i.e., the hamburger menu).  Tens of thousands of words have been written about the merits and more often the downsides of these two patterns (e.g. \cite{babich2017, jager2017, siang2015} for FAB and \cite{pernice2016, a2014, constine2014} for Hamburger Menu). Sometimes, the conclusions are daunting. For example, one online critic said, ``...in actual practice, widespread adoption of FABs might be detrimental to the overall UX of the app.''~\cite{siang2015}  Even when the criticisms are moderate and well-reasoned, they are based on the writer's examination of a limited number of examples. It is hard to know whether these criticisms reflect the full picture, since these patterns are likely to be used in a huge number of different apps. Thus the second research question driving this work is: \textit{How can we examine real world use of design patterns to inform debates about UI design?}

We took a big data approach to shed light on these two questions. We used two datasets in our analysis. The first dataset, \textit{Rico}~\cite{Rico}, consists of view hierarchies and screenshots of user interfaces from over 9,000 Android apps. The second dataset consists of app metadata from Google Play, the official app marketplace for Android. The metadata we used includes average app rating, number of installs, and category. Using text mining and computer vision techniques, we built a computational model to detect 6 widely-used UI components specified in Material Design, including the FAB, the Navigation Drawer and 4 other components. We then used the metadata of the apps in the first dataset to measure the relationship between use of certain patterns and the quality of apps, as indicated by app ratings and the number of installs. Furthermore, we used app category data to examine in what domain a certain pattern might be more useful. 

Our results show that use of Material Design is positively correlated with both the app's average rating and the number of installs, which we believe is the first quantitative evidence for the value of a pattern language applied to a large ecosystem in the wild. Our data analysis further shows that, despite the criticisms the FAB and the Navigation Drawer have received from vocal writers in the design community, they are more popular among apps with higher ratings and higher number of installs than their less popular peers. Furthermore, we found that use of UI components vary by app category, suggesting a more nuanced view needed in ongoing debates about UI design patterns.

\section{Related Work}
Our work is related to prior work in analyzing designs of mobile apps. Shirazi et al. mined UI layouts from 400 popular apps to understand user interface complexity and the popularity of common UI widgets (such as progress bars and checkboxes)~\cite{shirazi}. Alhrabi et al. mined and analyzed 24K Android apps to understand the usage of common design patterns (such as tabbed navigation)~\cite{alharbi}. These works predate the introduction of the Material Design specifications for Android and unlike our work, do not focus on understanding a pattern language nor its potential impact on app quality. Deka et al. mined mobile app design data at scale and demonstrated uses of such repositories for design applications~\cite{Rico, erica}. However, these works did not conduct any large scale design analysis of apps.

Outside of the mobile app design domain, our approach to understanding design using computational techniques is similar to works in multiple other domains. Kumar et al. created a web design repository with over 100K webpages and used computational methods to understand design demographics (such as popular colors and aspect ratios) in web pages~\cite{Kumar:2013}. Doosti et al. applied deep learning on the history of web design and tried to find patterns of each web design era beside find the influence of design pioneers on the other websites~\cite{Doosti:2017}. Reinecke et al. modeled visual preferences using user ratings on 430 website screenshots~\cite{Reinecke:2014}. O'Donnovan et al. developed an automated approach for learning good layout designs for graphic designs~\cite{odonnovan}. Although we take a similar overall approach to these works, ours is the first study that seeks to understand pattern language usage and its impact in the wild.

\section{Methodology}
There were two general stages in our data analysis. First, we detected Material Design elements in a large number of mobile apps. We focus on 6 elements, several of which are unique to Material Design: App Bar, Floating Action Button, Bottom Navigation, Navigation Drawer, Snack Bar, and Tab Layout. Second, we looked for relationships between usage of Material Design elements and app quality as well as app category. Below, we first describe the two datasets we used in this work, and then present the model we developed to automatically detect Material Design elements in Android apps.

\subsection{Datasets}
As mentioned earlier, we used the Rico dataset ~\cite{Rico} and the app metadata from Google Play. From the \textit{Rico} dataset, we analyzed over 72,000 Android UI screens. These UI screens were mined on Google Nexus 6P phones running the Android Marshmallow operating system from over 9,700 popular apps across 27 categories on the Google Play Store in early 2017. For each UI screen, the dataset contains both a screenshot as a \texttt{JPEG} file, and the hierarchical list of all the UI elements displayed on that screen (called a  \textit{View Hierarchy}) as a \texttt{JSON} file. For each UI element, the view hierarchy contains the classname, the names of superclasses, and its location in the screenshot, all of which were essential for developing our material design detectors that we describe in the next section.

From Google Play, we obtained the average rating, the number of installs, and app category for every app included in the \textit{Rico} dataset. Unlike app ratings, which are capped at 5, the number of installs does not have a maximum value and thus does not suffer from a potential ceiling effect. The apps included in the \textit{Rico} dataset have more downloads than apps in general. The median number of installs of apps in the \textit{Rico} dataset was about 1 million. 

\begin{figure*}
    \centering
    \includegraphics[width=\textwidth]{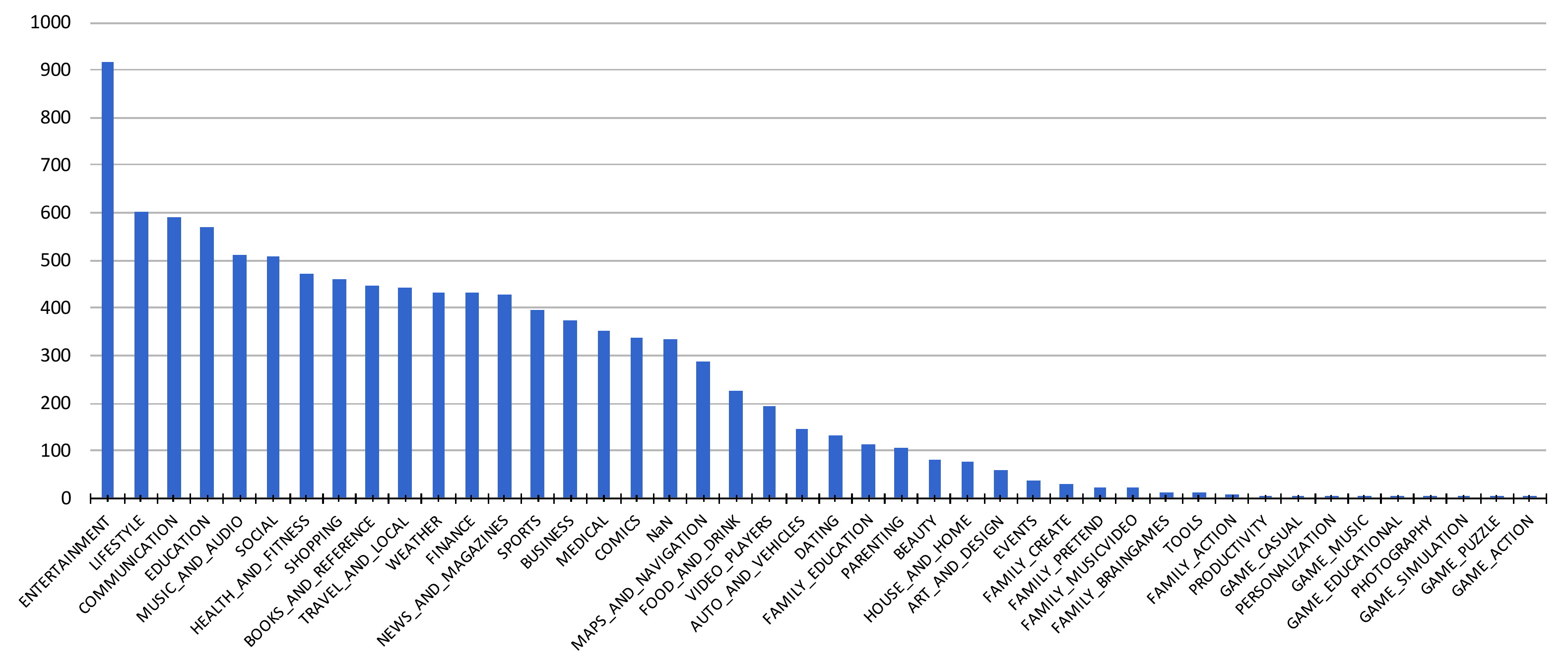}
    \caption{Category distribution of the apps in the Rico dataset}
    \label{fig:playdist}
\end{figure*}

Google Play Store categorizes apps in 58 different categories such as Education, Communication, Entertainment. 
The category distribution of the apps in the Rico dataset is shown in Figure~\ref{fig:playdist}.

\subsection{Detecting Material Design elements}

The main challenge in our data analysis was to reliably detect Material Design elements in apps. Our hope going into this work was that we could leverage standard implementations of Material Design elements to find their "signatures" in the view hierarchy of a UI, but we quickly found the variety of Material Design implementations to be too substantial to rely on matching known class names alone. Instead, we used a two phase approach. The first step for finding elements was to scan the view hierarchy files and look for strings that might match the name of the UI element. To accommodate different class naming practices in third-party and custom implementations of Material Design, we used very relaxed keywords to select samples. For example in the official Material Design library, the \textit{Floating Action Button} is implemented with a class named \texttt{FloatingActionButton}, but we used the keyword \texttt{Float} to capture more instances implemented in non-standard ways.

\begin{figure}[t]
\centering
\setlength\tabcolsep{1.5pt}
\begin{tabular}{ccccc}
\includegraphics[scale=0.17]{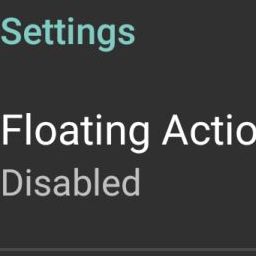} &
\includegraphics[scale=0.17]{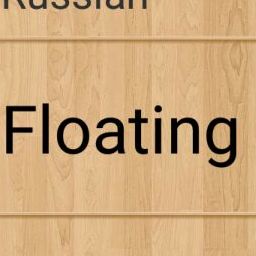} &
\includegraphics[scale=0.17]{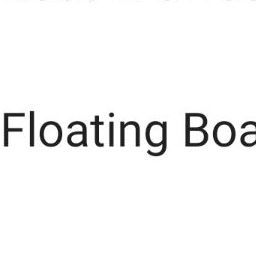} &
\includegraphics[scale=0.17]{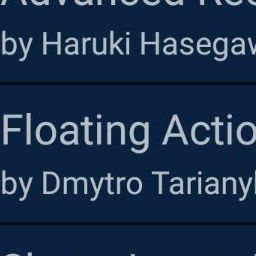} &
\includegraphics[scale=0.17]{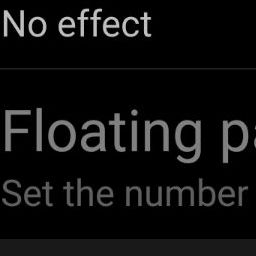}
\end{tabular}
\caption{Some false positives detected as Floating Action Button with keyword \texttt{float} which excluded in the Machine Intelligence step}
\label{fig:falsepositive}
\end{figure}

Although using relaxed keywords increased the detected elements, it could also increase the number of false positives. So we needed a complementary process to verify detected elements using their screenshots. We used computer vision techniques, especially the most common and robust computer vision tool called Convolutional Neural Network (CNN).
For each Material Design element we trained a network with the cropped images of official implementation of Material Design in the \textit{Rico} dataset and got at least 95\% of accuracy for all the elements.
Figure~\ref{fig:falsepositive} shows some of the false positives of the \textit{Floating Action Button} element with keyword \texttt{Float} which was excluded in the Machine Intelligence step.

\begin{figure*}[t]
\small
    \centering
    \begin{tabular}{ccccccc}
    \includegraphics[scale=0.45]{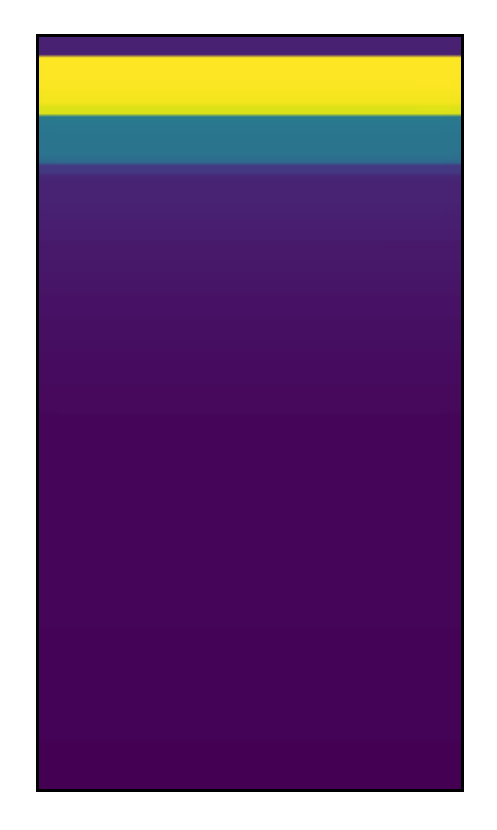} &
    \includegraphics[scale=0.45]{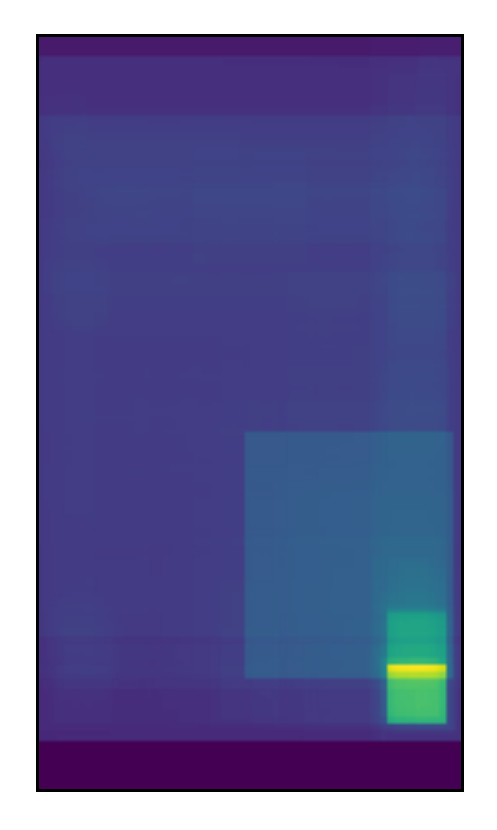} &
    \includegraphics[scale=0.45]{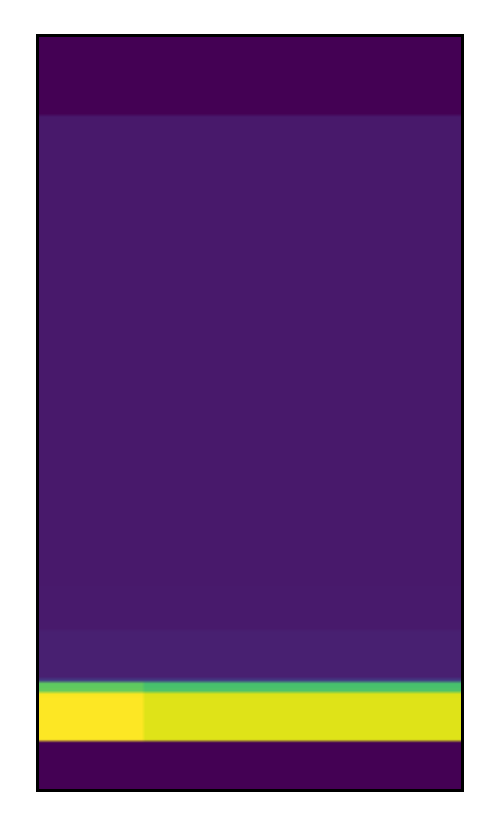} &
    \includegraphics[scale=0.45]{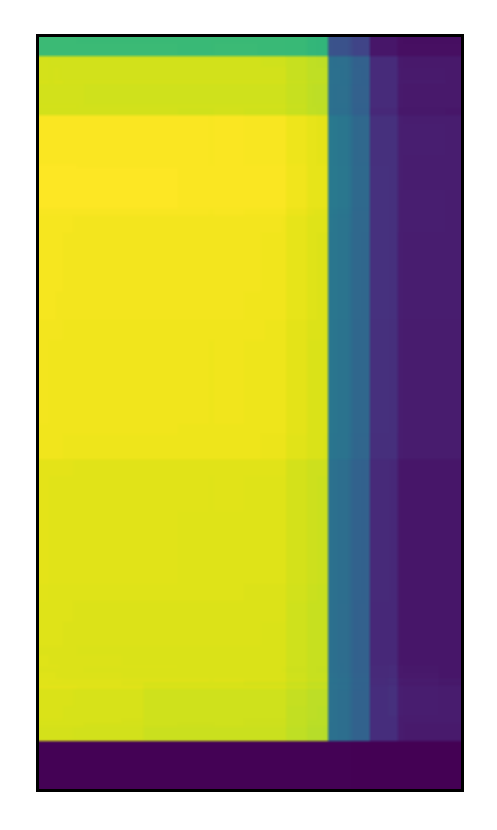} &
    \includegraphics[scale=0.45]{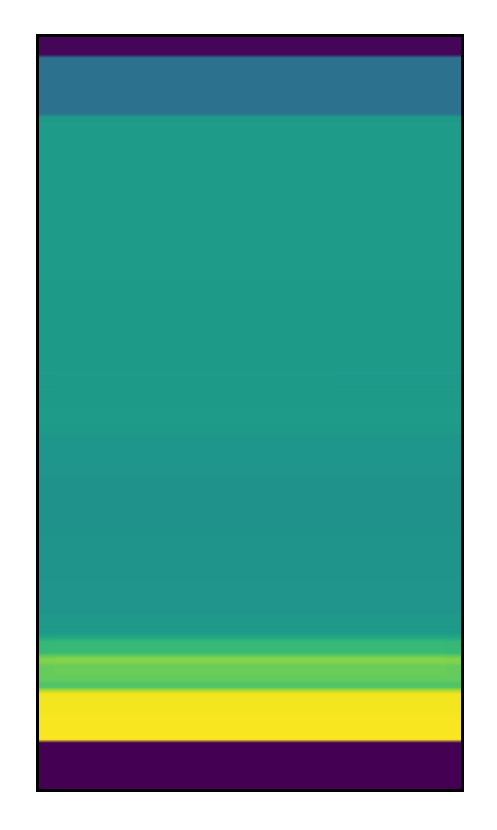} &
    \includegraphics[scale=0.45]{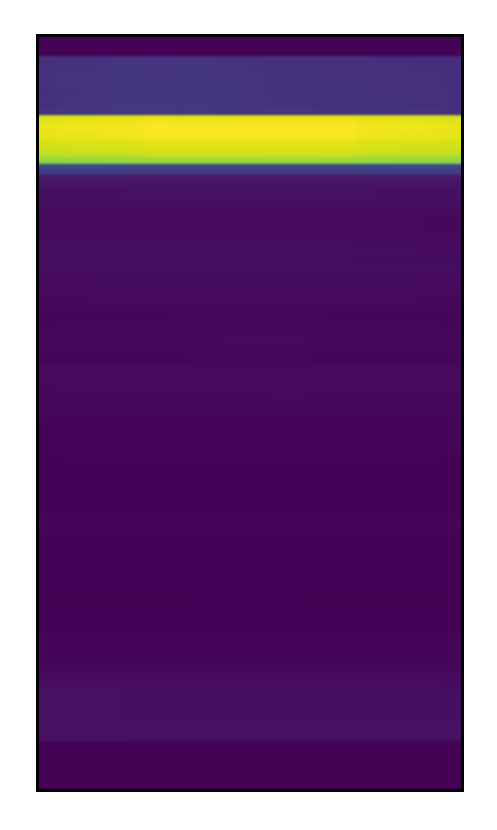} &
    \includegraphics[scale=0.45]{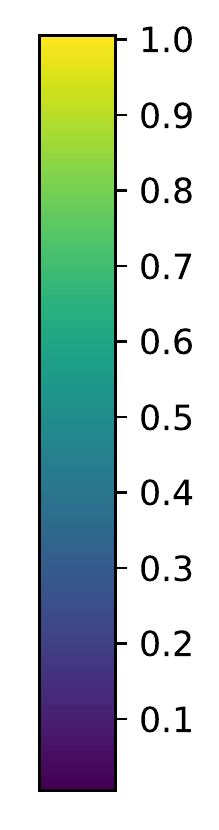}\\
    App Bar & Floating Action Button & Bottom Navigation &
    Navigation Drawer & Snack Bar & Tab Layout &
    \end{tabular}
    \caption{The heatmap of the frequency divided by maximum value of each Material Design element in \textit{Rico} dataset}
    \label{fig:heatmaps}
\end{figure*}

\subsubsection{Detecting Potential Elements from App View Hierarchy}
The \texttt{JSON} file in the \textit{Rico} dataset is a nested list of all the elements in that view. In this step we used five \texttt{JSON} keys which every component in the view hierarchy contained; \texttt{class}, \texttt{ancestors}, \texttt{bounds}, \texttt{visible-to-user} and \texttt{children} to find class name, super-classes names, location and visibility of element in the screen and the children of that element in that app view hierarchy respectively.
\texttt{class} is a string variable containing the name of the element.
The \texttt{ancestors} variable is a list of strings of all the names of the super class that object. If a developer inherits a class from the official library, the official class name will appear in the ancestors list.
The \texttt{bounds} variable is a list of integers indicating the coordinates of that element on the screen (Horizontal and Vertical coordinates of the top left and lower right of the element). We used \texttt{bounds} to locate the element on the screen and crop the screenshot image to be used in the computer vision part.
The \texttt{visible-to-user} variable is a boolean variable indicating the visibility of that item on the screen. We excluded items which were not visible on the screen.
The \texttt{children} variable is a list of other elements nested in that element.
Listing~\ref{samplejson} shows a sample of one element in the \texttt{JSON} file of the \textit{Rcio} dataset.

\begin{tiny}
\begin{lstlisting}[frame=none,caption=Sample of one element in the \texttt{JSON} file of the \textit{Rcio} dataset,label=samplejson]
{
  "resource-id": "se.perigee.android.seven:id/fab", 
  "adapter-view": false, 
  "pointer": "444477a", 
  "scrollable-horizontal": false, 
  "ancestors": [
    "android.support.design.widget.VisibilityAwareImageButton", 
    "android.widget.ImageButton", 
    "android.widget.ImageView", 
    "android.view.View", 
    "java.lang.Object"
  ], 
  "selected": false, 
  "content-desc": [null], 
  "rel-bounds": [1188, 1860, 1384, 2056], 
  "draw": false, 
  "focusable": true, 
  "long-clickable": false, 
  "visibility": "gone", 
  "focused": false, 
  "clickable": true, 
  "abs-pos": true, 
  "class": "android.support.design.widget.FloatingActionButton", 
  "visible-to-user": false, 
  "package": "se.perigee.android.seven", 
  "enabled": true, 
  "bounds": [1188, 2140, 1384, 2336], 
  "pressed": "not_pressed", 
  "scrollable-vertical": false
}
\end{lstlisting}
\end{tiny}

To find a Material Design element in the app view hierarchy, we had to look for the keyword corresponding to that element in the \texttt{class} variable. If the \texttt{class} value contained that keyword and it was visible to user we considered that app using that element. Since some users inherited from the official library class and made their own custom sub-class we checked the super-class names in the \texttt{ancestors} variable as well. If we did not find that keyword in the class name and super-class names we did the same operation on its children. Therefore the whole text mining process is a \textit{depth first search} process.

To remove false positives, we generated a collection of screenshots labeled with the detected UI elements in them. We used \texttt{bounds} variable to locate each element in the \texttt{JPEG} file and cropped that element with a margin. To build a classifer, we also needed some false samples for each component (e.g. a cropped part of the screenshot which is not a Floating Action Button) and for good performance, these samples should not be from a random place from the screen. Therefore, we started to record the places we find elements in the screenshots and made a frequency map for each element. Figure~\ref{fig:heatmaps} shows  heatmaps of normalized frequency for the 6 Material Design elements detected in \textit{Rico} dataset. Thus, when we could not find an element in a view, we cropped the most probable area of that element as a false sample. 

\subsubsection{Eliminating False Positives Using Screenshots}
We trained a separate classifier for each Material Design component 
based on the cropped images we collected in the previous step. We selected the AlexNet~\cite{alexnet} architecture for our Convolutional Neural Networks. As the task was relatively easy for a deep network, we trained all the networks from scratch with a learning rate of 0.001 and 50,000 iterations. We split our data to 80\% training, 10\% validation and 10\% for testing the trained network and got at least 95\% of accuracy on each component. We used Google's open-source machine learning platform TensorFlow~\cite{tensorflow} for all our machine learning and deep learning computations.

Note that there is a probability that one app used one Material Design component but it is not visible in the cropped image. There were lots of views in which Floating Action Button was used but it was occluded by Android keyboard or App Bar was covered by Navigation Drawer. To prevent removing these true positive samples, CNN checked every cropped image of that app and marked that app did not use a particular element if it did not find that element in all the cropped images of that app. 

\begin{figure}[t]
    \centering
    \begin{tabular}{cc}
        \includegraphics[width=0.15\textwidth]{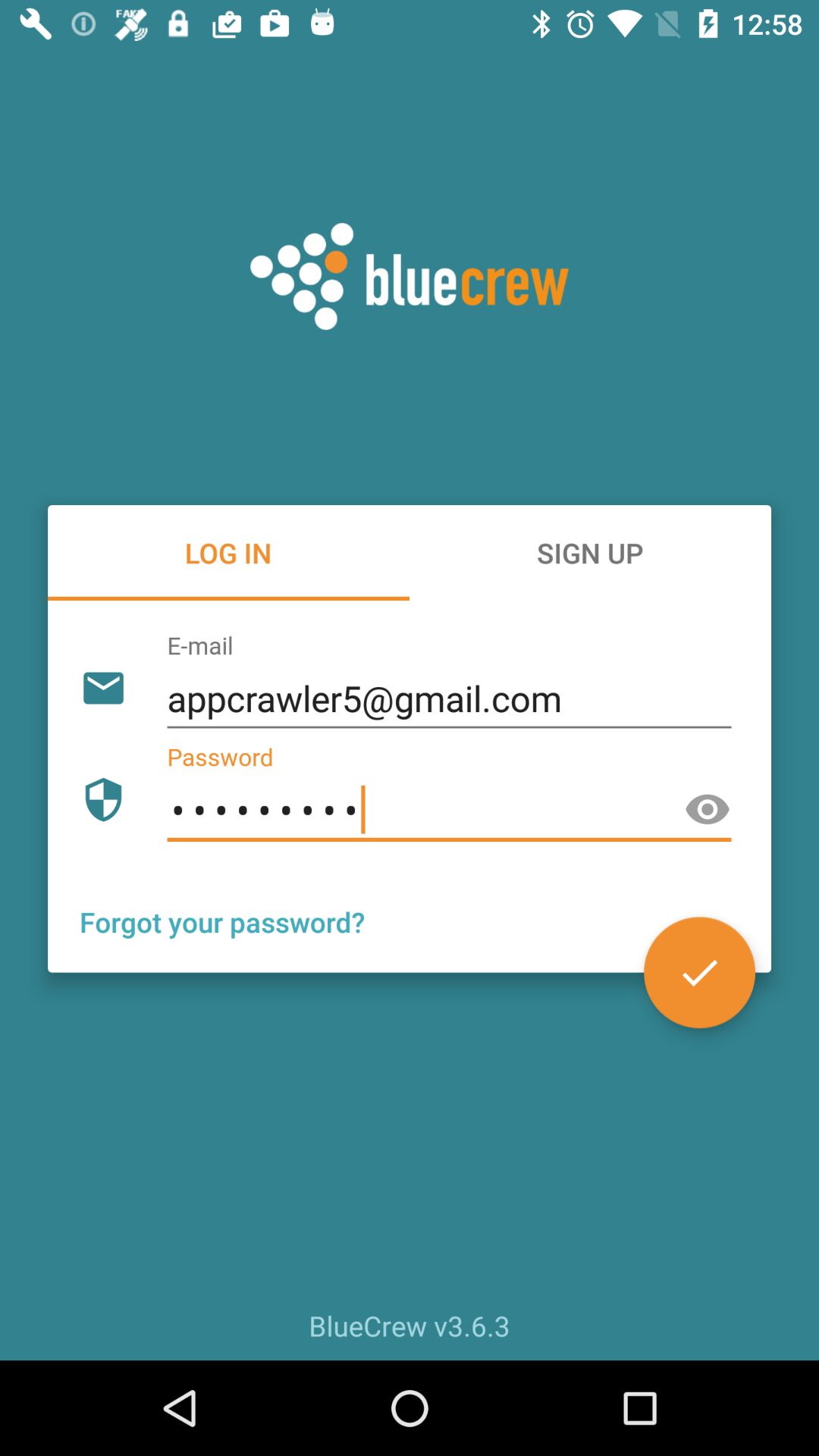} & \includegraphics[width=0.15\textwidth]{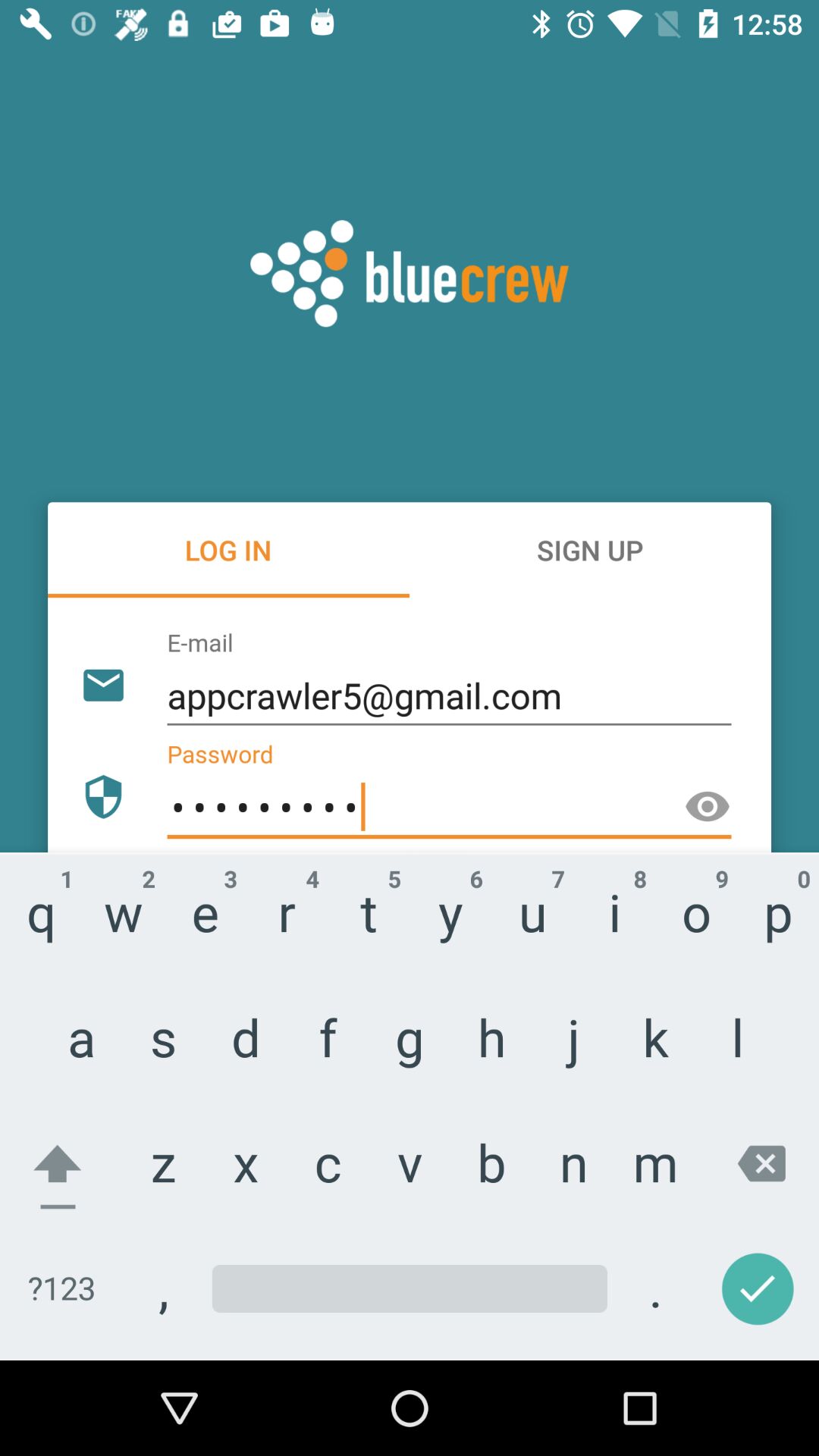}
    \end{tabular}
    \caption{Floating Action Button (left) and that component occluded by Android keyboard (right) in two different views of an app in \textit{Rico} dataset}
    \label{fig:occluded}
\end{figure}

Figure~\ref{fig:occluded} shows two screenshots of an app in the \textit{Rico} dataset. You can see that the orange Floating Action Button is occluded by Android keyboard in one of the views.

\section{Results}
In this section, we first report the usage of the Floating Action Button and the usage of the Navigation Drawer, two frequently criticized patterns in Material Design, in relation to the average app rating, the number of installs, and the app category. We then report the usage of Material Design in general and examine its impact on app quality.

\subsection{Usage of Floating Action Buttons}
If the drawbacks of FABs generally outweigh their benefits, one would assume that developers who build higher-rated apps would know better and are less likely to use FABs in their apps than those who build lower-rated apps. To test this hypothesis, we split apps into two groups: a high-rating group and a low-rating group by the median average rating of all apps in the \textit{Rico} dataset, which was 4.16. In other words, apps in the low-rating group had average ratings lower than 4.16, while those in the high-rating group had average ratings higher or equal to 4.16. The two groups of apps are balanced and each group have 4673 number of apps.

\begin{figure*}[t]
    \centering
    \begin{tabular}{cccc}
    \includegraphics[scale=0.7]{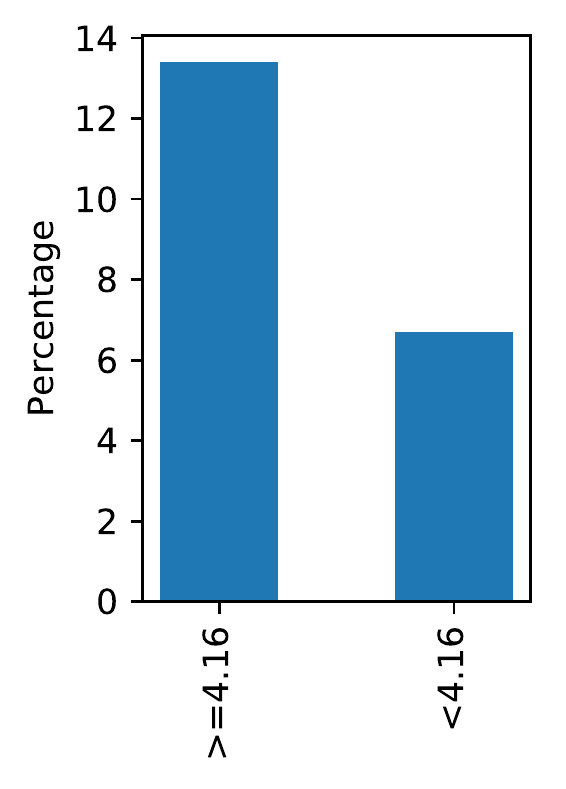} &
    \includegraphics[scale=0.5]{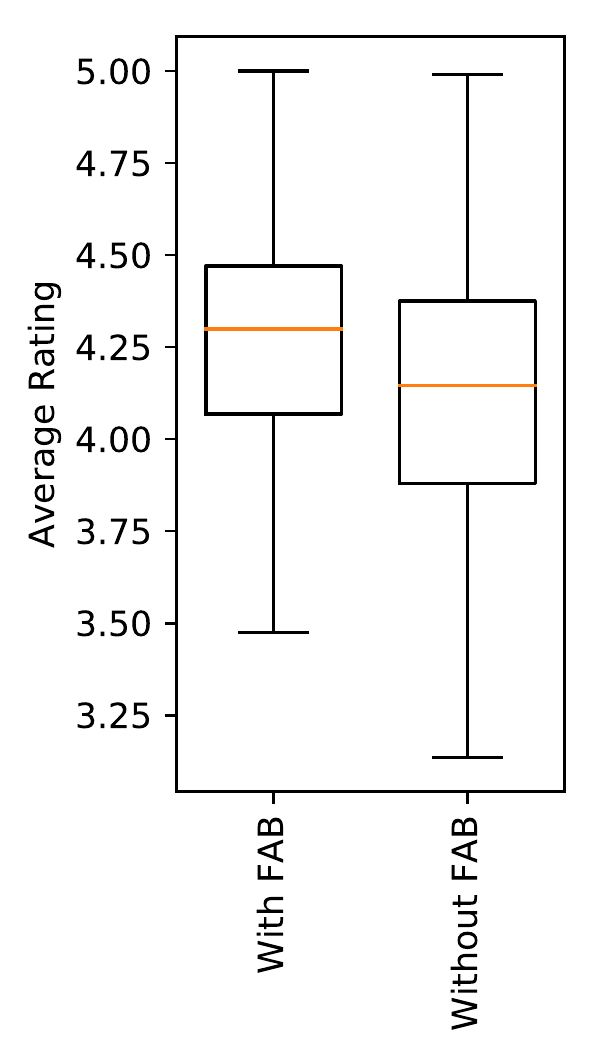} &
    \includegraphics[scale=0.7]{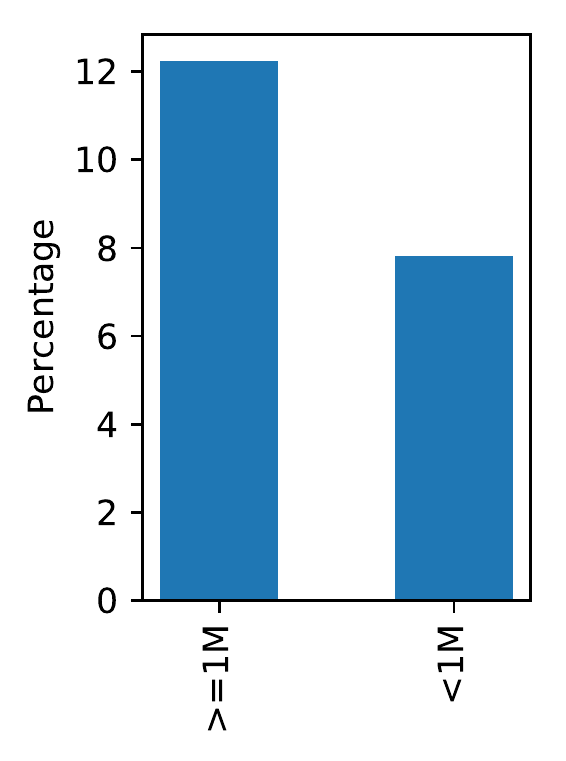} &
    \includegraphics[scale=0.5]{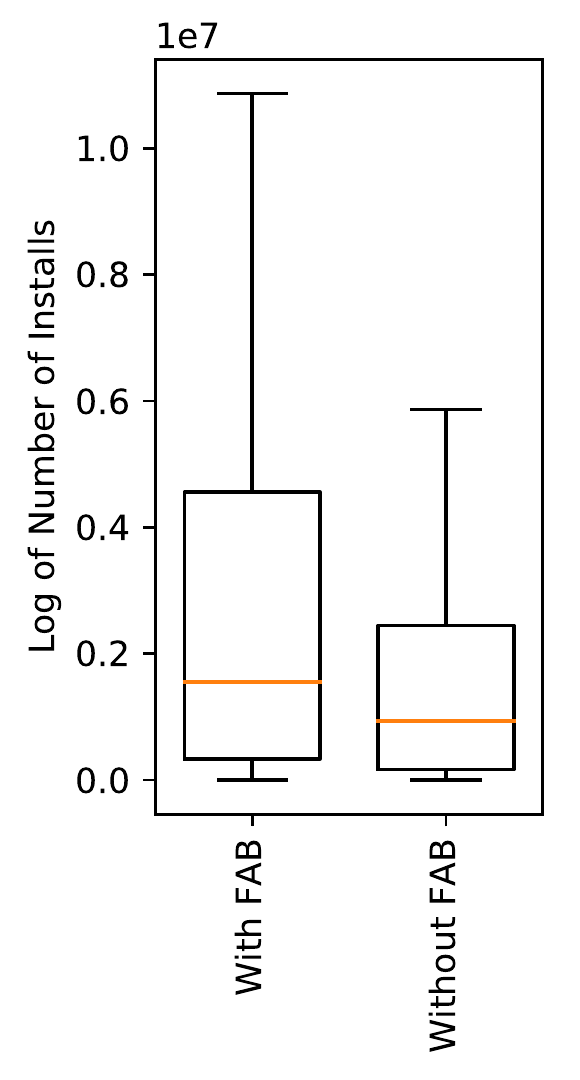} \\
    (a) & (b) & (c) & (d)
    \end{tabular}
    \caption{(a) The percentage of apps using the Floating Action Button (FAB) in the high-rating group vs. the low-rating group (b) Box plots of the average ratings of apps using the FAB vs. those not using the FAB (c) The percentage of apps using the FAB in the more-installed group vs. the less-installed group (d) Box plots of the number of installs of apps using the FAB vs. those not using the FAB}
    \label{fig:float}
\end{figure*}

As Figure~\ref{fig:float}.a shows, to our surprise there are a higher percentage of apps using FABs in the high-rating group than those in the low-rating group (13.4\% vs 6.6\%). The box plots in Figure~\ref{fig:float}.b further show that  apps using the FAB were rated higher than those did not use it. In fact, 66.6\% of apps that used the FAB belonged to the higher-rating group.

We also used the number of installs as another measure of app quality. Thus, we decided to split our apps into two groups: 1) apps with greater than or equal to 1 millions of installs, and 2) apps with less than 1 millions of installs. The two groups were nearly balanced after the split, with 4723 in the more-installed group and 4623 in the less-installed group). Similar to what we saw in the analysis of FAB usage and app ratings, apps in the more-installed group appeared to be more likely to feature FABs than those in the less-installed group (see Figure~\ref{fig:float}.c). Also apps using the FAB have more number of installs in comparison to apps without the FAB (see Figure~\ref{fig:float}.d).

The results above suggest that many developers of high-quality apps consider the FAB to be a valuable design pattern. Although this would seem to invalidate some of the controversy around the FAB, it is not conclusive. It may be, for example, that the FAB is a more usable pattern in some situations than others. 

\begin{figure}[t]
    \centering
    \includegraphics[scale=0.7]{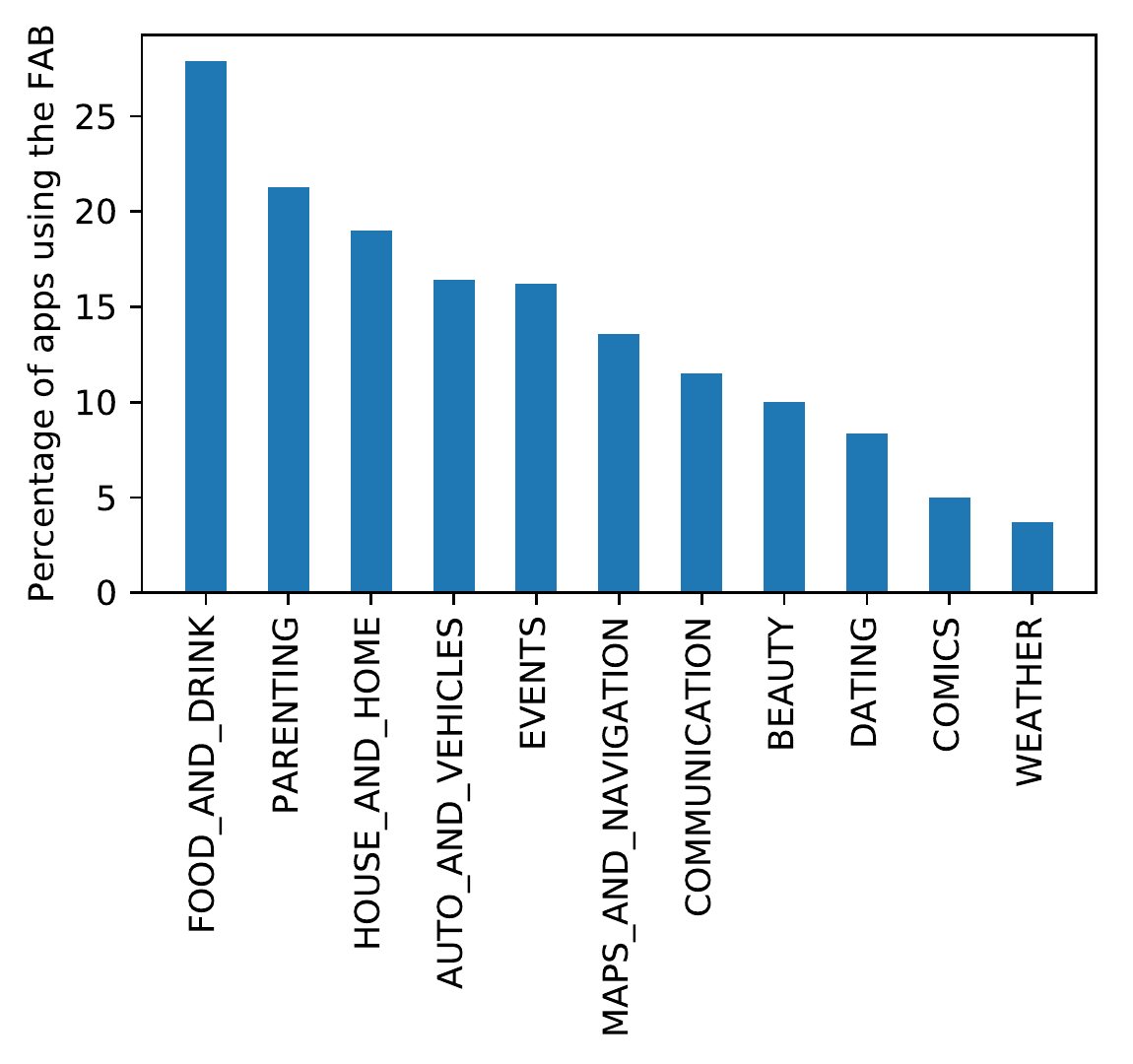}
    \caption{The top category of apps which used the most FAB by percentage in their category}
    \label{fig:floatcategory2}
\end{figure}

\begin{figure*}[t]
    \centering
    \includegraphics[width=\textwidth]{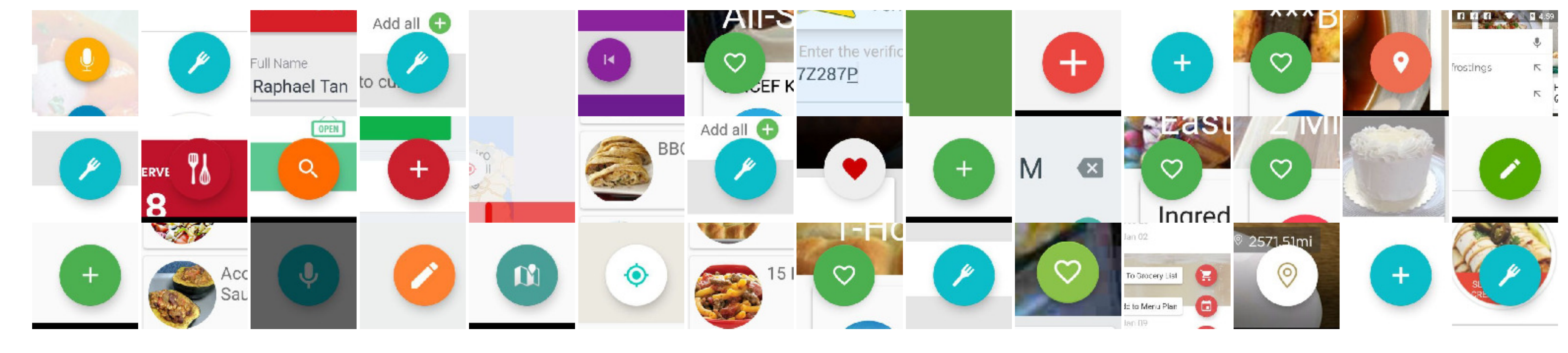}
    \caption{Thumbnails of FABs in the Food and Drink category apps}
    \label{fig:floatcollage}
\end{figure*}

To understand where the FAB might be more useful, we examined usage of the FAB by app category. Figure~\ref{fig:floatcategory2} shows the top 11 app categories by the percentage of apps featuring the FAB, excluding categories for which there were too few apps in the \textit{Rico} dataset (less than 0.05\% of the apps of that category in Google Play). As it is obvious to see, FAB usage varied considerably among these 11 categories of apps. The \textit{Food and Drink} category had the highest percentage of FAB usage among all the qualified categories. Figure~\ref{fig:floatcollage} shows some of the FABs in the \textit{Food and Drink} category. Each thumbnail belongs to a different app but you can see that there are similar thumbnails in this category. For example the \textit{fork} icons belongs to six different apps that are all developed by \textit{Riafy Technologies}, a producer of recipe apps for special diets like diabetes. Also there are some food delivery apps in this category which used the FAB to point at the current location in the map. Note that some of the thumbnails in this picture do not appear to include a FAB, because they are occluded by another UI component.

\subsection{Usage of Navigation Drawers}
\begin{figure*}[t]
    \centering
    \begin{tabular}{cccc}
    \includegraphics[scale=0.7]{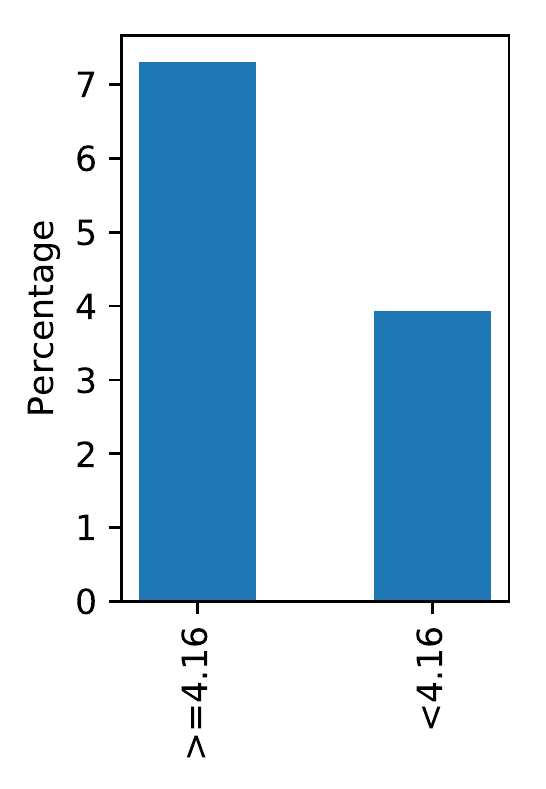} &
    \includegraphics[scale=0.5]{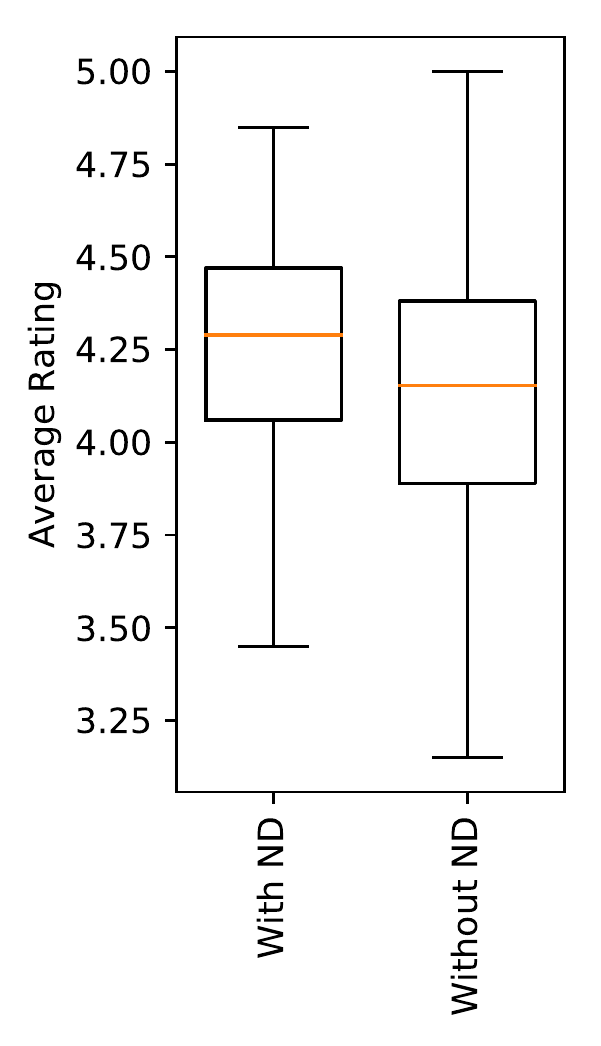} &
    \includegraphics[scale=0.7]{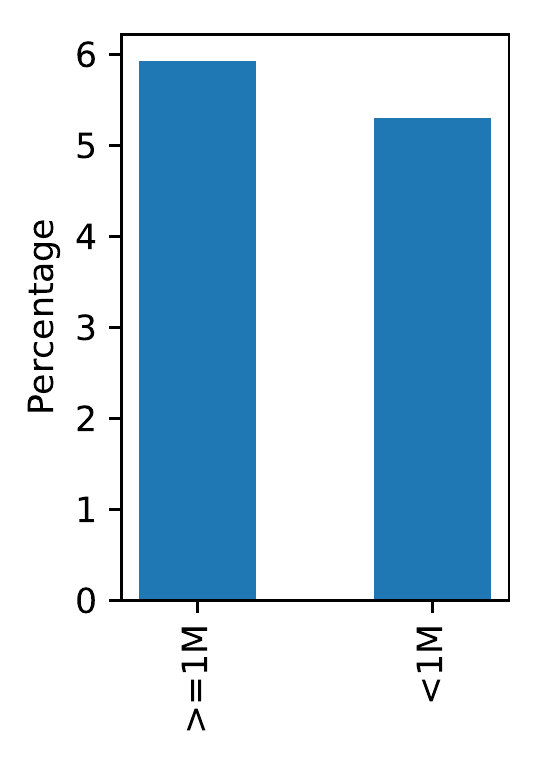} &
    \includegraphics[scale=0.5]{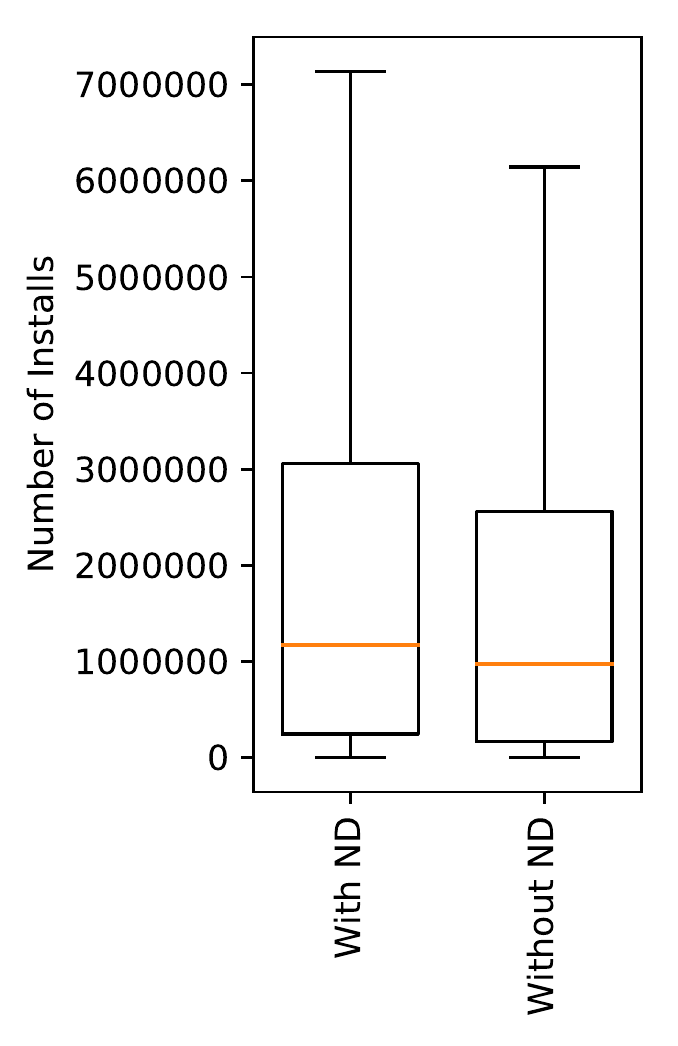} \\
    (a) & (b) & (c) & (d)
    \end{tabular}
    \caption{(a) The percentage of apps using the Navigation Drawer in the high-rating group vs. the low-rating group (b) Box plots of the average ratings of apps using the Navigation Drawer vs. those not using the Navigation Drawer (c) The percentage of apps using the Navigation Drawer in the more-installed group vs. the less-installed groups (d) Box plots of the number of installs of apps using the Navigation Drawer vs. those not using the Navigation Drawer}
    \label{fig:nd}
\end{figure*}

After examining the usage of the FAB, we then applied the same analysis to the Navigation Drawer, another controversial pattern in Material Design. As we can see in Figure~\ref{fig:nd}.a, there were more apps in the high-rating group which had a Navigation Drawer than those in the low-rating group (7.3\% vs. 3.9\%). The box plots in Figure~\ref{fig:nd}.b show that the average rating for apps using the Navigation Drawer was higher than those that did not use it. Among all the apps that used the Navigation Drawer, 65\% of them belonged to the high-rating group.

Similar to our analysis of the FAB usage, we examined the usage of the Navigation Drawer and the number of installs. As it is shown in Figure~\ref{fig:nd}.c apps in the high-rating group were a little more likely to feature a Navigation Drawer than those in the low-rating group. Also, the box plots in Figure~\ref{fig:nd}.d show that apps using Navigation Drawer had slightly higher number of installs.

\subsection{Material Design and App Quality}
We conducted an analysis to understand the usage of Material Design in general and its relationship to apps' average ratings and number of installs. The first step was to determine if an app used Material Design. We adopted a relatively relaxed criterion: if an app used one of the six Material Design components our design analyzer could detect, we considered that app as using Material Design.

\begin{figure}[t]
    \centering
    \includegraphics[scale=0.6]{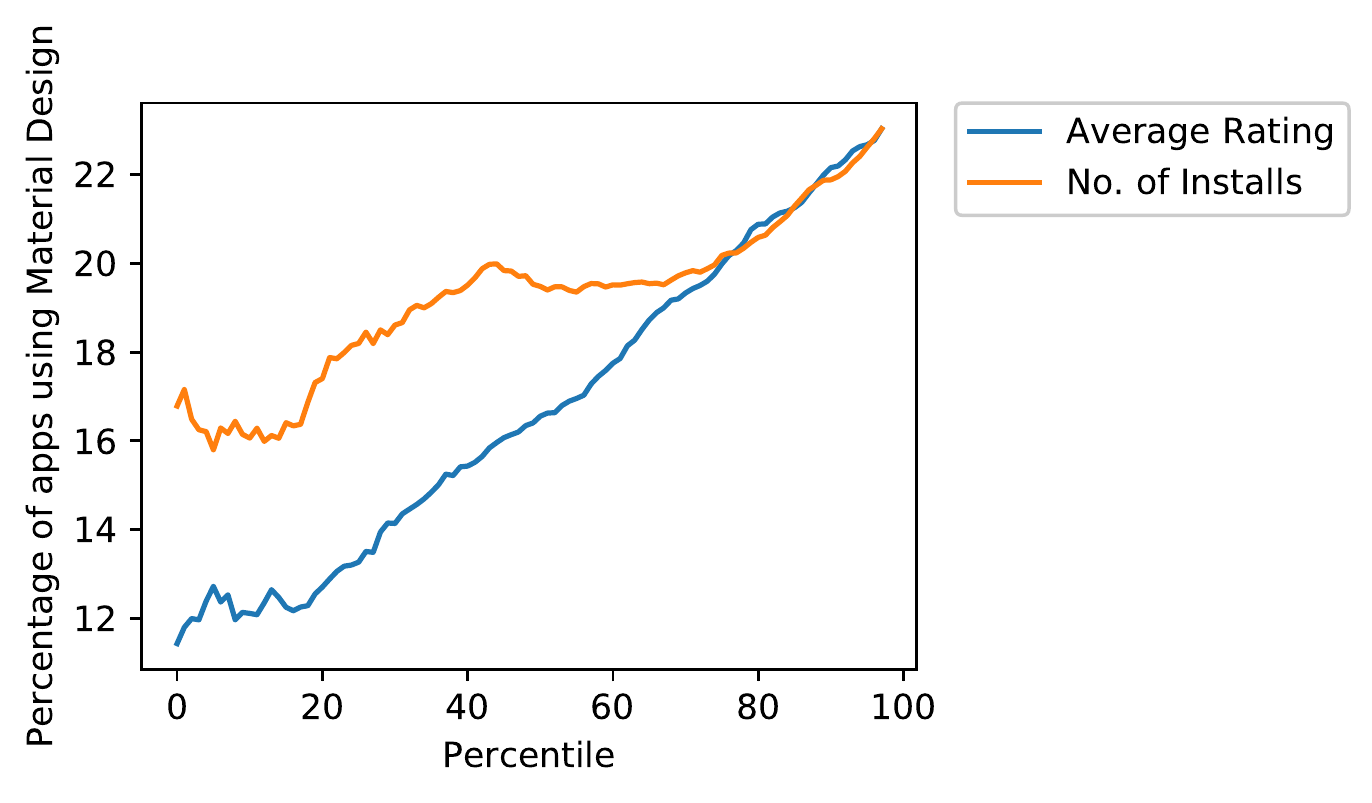}
    \caption{Distribution of the percentage of apps using at least one of the six common Material Design elements over percentiles in average rating (blue) and number of installs (orange)}
    \label{fig:mdmix}
\end{figure}

First, we examined the relationship between the usage of Material Design and apps' average ratings. To this end, we sorted all the apps in the \textit{Rico} dataset by their average ratings, split them into one hundred buckets and calculated the percentage of apps that used Material Design for each percentile. We then plotted the percentage of Material Design usage over average rating percentile. As we can see in Figure~\ref{fig:mdmix}, the usage of Material Design was highly correlated with the average rating percentile (with Pearson correlation coefficient \(\rho = 0.99\) and p-value = \(3.1 \times 10^{-91}\)). In other words, as the average rating increased, the usage of Material Design also increased.

Next, we examined the relationship of the usage of Material Design and the number of installs, an alternate measure of app quality. As in the previous step, we sorted apps by their number of installs and split them into one hundred equal-sized buckets by percentile. As shown in Figure~\ref{fig:mdmix}, the percent of the apps using Material Design is also highly correlated to number of installs \(\rho = 0.94\) and p-value = \(2.3 \times 10^{-47}\)).

\section{Discussion}
In this work, we explored a methodology to enhance our understanding of UI pattern language usage in the wild and applied that methodology to the Material Design pattern language. Here, we review the two research questions we set out to answer, discuss the limitations of our analysis, and suggest future work.

Our first research question was \textit{How can we understand the impact of a pattern language in one of the largest computing ecosystems in the world?} We developed a computational method to measure the relationship between Material Design, the pattern language in question, and app quality in the Android ecosystem. The method involves two stpdf. First, we used data mining techniques to detect UI design patterns in app view hierarchy data and removed false positives using Computer Vision techniques applied to app screenshots. Second, we combined the pattern usage data and signals of app quality from the app metadata obtained from Google Play to investigate potential relationships between the two. Our analysis has shown that both app ratings and number of installs were positively correlated with Material Design usage for the apps in the \textit{Rico} dataset. This result suggests that well-crafted pattern languages can help raise the overall user experience of a large app ecosystem.  

Our second research question was \textit{How can we examine real world use of design patterns to inform debates about UI design?} To answer this question, we examined usage of the Floating Action Button and the Navigation Drawer, two frequently criticized patterns in online design discussions. While our results do not directly rebut specific arguments against these two patterns, they clearly show that many developers and designers found these two patterns valuable and use both patterns frequently in their highly-rated and highly-popular apps. Moreover, our results have suggested that evaluating the merit of a design pattern should consider the context it is applied to. For example, developers used the FAB more frequently in certain app categories such as \textit{Food and Drink} and \textit{Parenting} than others. Our preliminary visual inspection of apps in these top categories found that many of those apps allowed the user to take a clear primary action on their user interfaces where a FAB was used, suggesting a match between user tasks and design patterns.

\subsection{Limitations}
We acknowledge that our analysis is limited by the data available to us. First, the number of apps included in the \textit{Rico} dataset is small compared to the total number of apps in the entire Play Store. This limitation is partially mitigated by the fact that the \textit{Rico} dataset includes more popular apps, which ought to bear a heavier weight with our analysis on the impact of pattern languages than less-used apps. Second, there is a group of apps in the \textit{Rico} dataset which the crawler could not pass the login screen, so we had to exclude 320 apps or 3.4\% of total apps in \textit{Rico} in our analysis. Since there is no good reason to believe apps that required login might be more or less likely to use certain design patterns, excluding such apps should not have changed the findings from our analysis.

\subsection{Future work}
There are many ways we would like to extend this work. First, it would be useful to track usage of design patterns over the time and correlate those changes to app ratings and number of installs. To achieve this, we will need to run the crawler used by \textit{Rico} over multiple versions of the same app and develop an infrastructure to save historic data from both the app crawler and the Play Store. Second, we would like to extend our analysis to analyze apps from different geographical regions, since research has shown differences in visual preferences across cultures that might affect design language usage~\cite{Reinecke:2014}. Last, we can refine the way we looked at pattern usage and its context. In addition to using app category, we could potentially crowdsource task labels for each app view and examine how UI patterns and tasks might be associated with one another.

\section{Conclusion}
In conclusion, we developed a computational method to examine the use of UI design pattern languages in the wild by combining app hierarchy, screenshots, and marketplace metadata. Our data analysis showed that this method was effective in demonstrating the overall impact of a pattern language, Material Design, on the quality of apps in the Android ecosystem. Furthermore, leveraging a large-scale design dataset can shed light on debates about the merits of popular design patterns such as the Floating Action Button and the Navigation Drawer and provide context often missing from such debates.


\bibliographystyle{ACM-Reference-Format}
\bibliography{sample-sigconf}

\end{document}